\documentclass[prd,twocolumn,superscriptaddress,showpacs,amsmath,amssymb]{revtex4}
\usepackage{graphicx}
\usepackage{amsmath}
\usepackage{amssymb}
\usepackage{float}
\usepackage{color}
\usepackage{braket}

\usepackage{cancel,xcolor}
\usepackage{indentfirst}
\usepackage{CJKulem}
\usepackage{soul,xcolor}
\setstcolor{red}
\usepackage{ulem}
\usepackage{cancel}
\usepackage[pdfpagemode=UseNone,pdfstartview=FitH,colorlinks=true,linkcolor=blue,urlcolor=blue,anchorcolor=blue,citecolor=blue]{hyperref}

\begin{document}

\title{\titlename}
\date{\today}
\title{Krylov complexity for $\eta$ deformed superstring backgrounds}
\author{Dibakar Roychowdhury}
\email{dibakar.roychowdhury@ph.iitr.ac.in}
\affiliation{Department of Physics, Indian Institute of Technology Roorkee
Roorkee 247667, Uttarakhand,
India}

%\date{\today}
\begin{abstract}
We compute the holographic Krylov complexity for a class of strongly coupled Quantum Field Theory (QFT) (with broken scale invariance) in a top down approach, where the dual gravitational counterpart corresponds to $\eta$ deformed superstring solutions in a type IIB set up. The full 10d solution contains the anti-de Sitter ($AdS$) as the subspace, along with the dilaton and the background Ramond-Ramond (RR) and Neveu-Schwarz (NS) fluxes. The Krylov complexity in the dual operator picture is obtained by computing the proper momentum of the massive particle along its geodesic in the bulk spacetime. We explore the effects of $\eta$ deformation on the bulk momentum of the particle, which in turn affects the rate of growth of complexity in the dual QFTs. Our results boil down into the undeformed case in the appropriate limit of the Yang-Baxter (YB) parameter. We also comment on the complexity of the dual quantum mechanical model with broken scale invariance.
\end{abstract}

\maketitle
%%%%%%%%%%%%%%%%%%%%%%%%%%%%%%%%%%%%%%%%%%%%%%%%
\section{Introduction and General Idea}
Chaotic systems can be diagnosed by means of out-of-time-order correlators (OTOCs) \cite{Maldacena:2015waa}, which exhibit exponential behavior characterized by the Lyapunov exponent. Another quantity that has recently drawn renewed attention in the context of quantum chaos is the Krylov complexity \cite{Parker:2018yvk}-\cite{Anegawa:2024wov} and is the latest addition to existing models on complexity \cite{Stanford:2014jda}-\cite{Belin:2021bga} in the literature. The Krylov complexity typically is the measure of how quickly an operator $\mathcal{O}$ spreads over the Krylov subspace of the full Hilbert space under time evolution \cite{Caputa:2025dep}-\cite{Balasubramanian:2025xkj}. This might exhibit oscillatory, linear, or exponential growth depending on the nature of the system under consideration.

In this work, we shall be mostly aligned with the arguments of \cite{Susskind:2018tei}-\cite{Ageev:2018msv} which conjecture that operator-size growth in holographic QFTs corresponds to the growth of the radial momentum of a particle in the dual gravitational description. We apply this idea to various deformed supergravity backgrounds and conjecture about the Krylov complexity associated with the dual-operator growth. According to conjecture \cite{Susskind:2018tei}-\cite{Ageev:2018msv}, the rate of complexity growth is proportional to the proper radial momentum of the massive particle in the bulk spacetime
\begin{align}
\label{e1.1}
    \partial_t\mathcal{C}(t) =-\frac{P_\rho}{\epsilon}.
\end{align}

The particle (of mass $m$) is created near the boundary of the spacetime due to the insertion of an operator $\mathcal{O}(t_0,x)$ at some particular instant of time ($t_0$). As time progresses, the size of the operator increases $\mathcal{O}(t,x)=e^{i H t}\mathcal{O}(0,x)e^{-iH t}$, resulting in a cascade of commutators and, therefore, in an increase in complexity ($\mathcal{C}(t)$). In the dual gravity description, the particle starts to fall through the bulk without back-reacting on the geometry.

There have been two different approaches to the Krylov complexity in the context of holography. One of them is based on the above theme of computing the proper momentum of the particle along the geodesic \cite{Balasubramanian:2022tpr}-\cite{He:2024pox}, while the other is based on the notion of the rate of growth of the wormhole length in the dual gravity description \cite{Heller:2024ldz}-\cite{Ambrosini:2024sre}. The reader is also encouraged to refer to \cite{Baiguera:2025dkc}-\cite{Rabinovici:2025otw} for a nice set of comprehensive reviews on the subject.

As emphasized above, in this paper, we focus solely on the Krylov complexity \cite{Parker:2018yvk} growth associated with the operator spectrum in dual QFTs following the former line of approach. In particular, we extend the arguments of \cite{Caputa:2024sux} to compute the Krylov complexity growth for a class of Yang-Baxter (YB) deformed supergravity backgrounds \cite{Hoare:2018ngg}-\cite{Hoare:2014pna}. 

\paragraph*{Motivation and Lessons Learned.} 
Extending the previous analysis of the holographic Krylov complexity \cite{Caputa:2024sux} to Yang-Baxter (YB) deformed backgrounds is particularly interesting and compelling from various perspectives. First, it provides a controlled top-down set up to explore how integrable or field-theoretic deformation parameters directly govern operator growth dynamics and quantum chaos in dual QFTs with broken scale invariance \cite{Roychowdhury:2017oqd}. Studying these integrable deformations further allows us to explore how the explicit breaking of bulk isometry is reflected on the underlying Krylov basis and the corresponding Krylov dynamics of operator spreading. As we show below, the central lesson emerging from our analysis is that the YB deformation parameter $\kappa$ effectively introduces a finite radial cut-off in the bulk geometry that is manifested as a holographic screen at finite radial distance $\rho_B \sim \sinh^{-1}(1/\kappa)$ where the dual QFT is living. In the language of the field theory, the presence of the holographic screen eventually truncates the effective dimensionality of the dual Krylov basis. Physically, YB deformation acts as a damping mechanism: increasing the parameter value eventually suppresses the rate of operator complexity growth in the dual field theory compared to its undeformed CFT parent, offering a concrete bulk geometric mechanism for controlling chaos and operator dynamics in deformed holographic theories.

We begin by discussing the Yang-Baxter (YB) deformed $AdS_2 \times S^2 \times T^6$ in 10d \cite{Hoare:2018ngg}-\cite{Roychowdhury:2025aye}. These are the solutions in type IIB supergravity constructed using unimodular R-matrices. We show that these YB deformed geometries are conformally equivalent to the conventional Krylov basis \cite{Caputa:2024sux} where the conformal/scale factor explicitly depends on the YB deformation parameter. To simplify calculations, we consider a particular geodesic where the background dilaton becomes a trivial factor; as a result the string and the Einstein frame are the same. 

In the subsequent analysis, we generalize the above result by uplifting the QFT in one higher dimension whose dual gravitational counterpart corresponds to a supergravity solution with a pure $AdS_3$ factor plus deformations due to YB effects. These backgrounds are generated through unimodular $R$-matrices that depend on the deformation parameters ($\kappa_{\pm}$) associated with each copy of the superisometry algebra of the undeformed $AdS_3 \times S^3 \times T^4$ \cite{Hoare:2014oua}-\cite{Seibold:2019dvf}. Depending on the choice of the $R$-matrix (with all fermionic roots), two different supergravity solutions are obtained. One of these deformed backgrounds allows for a mirror dual, which we also explore towards the end of this paper. As these backgrounds are manifested as deformations of the standard $AdS_3$ background, they bring up some non-trivial modification to the standard $SL(2,R)$ Krylov spin chain \cite{Caputa:2024sux}. These modifications on the field theory side are in one to one correspondence with the YB deformation in the bulk, and their effects can be seen in a non-trivial fashion within the holographic setup.

\begin{figure}
    \centering
    \includegraphics[width=0.35\textwidth]{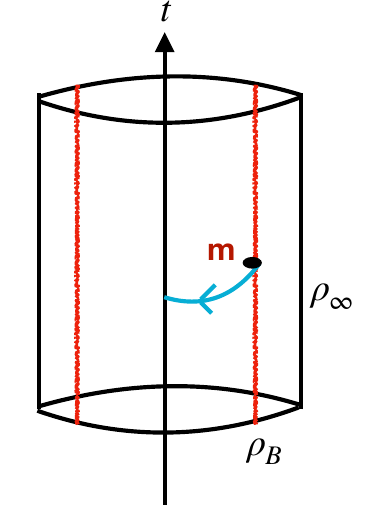}
    \caption{In the Figure above we picturize the geodesic motion of a massive particle (of mass $m$) in the Yang-Baxter (YB) deformed background in global coordinates. The ``holographic screen'' (which could be thought of as an effective boundary of the spacetime) is located at finite radial distance $\rho_B$, which acts like a radial cut-off in the bulk AdS. Here, $\rho_\infty$ is the location of the true asymptotic boundary of the (undeformed) AdS. From the perspective of an observer sitting at $\rho_\infty$, the massive particle is created at $\rho_\Lambda \sim \rho_B$, which is therefore always at a finite radial distance from the true boundary of the AdS spacetime and is identified with a finite momentum to begin with. The YB deformation reduces the proper geodesic distance of the massive particle trajectory in the bulk and hence the rate of complexity growth for the dual QFT living on $\rho_B$.}
    \label{figYB}
\end{figure}

The purpose of our analysis is in particular to explore the effects of YB deformation on the Krylov complexity growth of operators in dual QFTs. To see this explicitly, we compute the proper momentum ($P_\rho$) of the particle as a function of time ($t$) and calculate its variation for different choices of the YB parameter. In our first example of deformed $AdS_2 \times S^2 \times T^6$, we see that the effects due to YB deformations can be interpreted from the perspective of the undeformed theory as a particle moving with a finite initial velocity near the boundary, which corresponds to a non-zero rate of growth of complexity to begin with (see Fig.\ref{figYB}). The effect of YB deformation is to introduce a finite radial cut-off in the bulk, which reduces the effective dimensionality of the Krylov basis in the dual CFT. Almost identical features are observed for the two parameter deformation of the $AdS_3 \times S^3 \times T^4$ background \cite{Seibold:2019dvf}, except for the non-trivial fact that in this particular example, one can perform analytic calculations with a very careful choice of deformation parameters ($\kappa_L$ and $\kappa_R$) setting both of them non-zero and very close to each other. It turns out that in this particular limit $\kappa_L \sim \kappa_R$, the geodesic of the massive particle in the dual geometry probes an $SL(2,R)$ sector, which is therefore identified as the Krylov basis \cite{Caputa:2024sux} in the dual QFT.

The mirror model is indeed quite different from its parent theory. To identify the Krylov basis, one has to perform a double Wick rotation and thereby an expansion in coordinates. It appears that the background has an exact $SL(2,R)$ symmetry for zero deformation and corresponds to a 2d CFT on a thermal cycle with periodicity $\beta=2\pi$. However, in the presence of finite YB effects, one deviates from this ideal situation. In all three cases, we find a suppression of the rate of complexity growth with increasing YB deformation. 

The organization of the rest of the paper is as follows. We begin by reviewing the fundamentals of Krylov operator growth in Section \ref{sec:krylov_intro}. In Section \ref{sec3}, we identify the dual Krylov basis and the associated complexity for $\eta$ deformed $ AdS_2 \times S^2 \times T^6 $ solution in type IIB supergravity. We extend them for two parameter deformed $AdS_3 \times S^3 \times T^4$ supergravity solution in the next Section \ref{sec4}, where we identify the appropriate Krylov basis in a particular limit of the deformation parameters. In the following Section \ref{sec5}, we consider the mirror model and show that the dual Krylov basis can be obtained following a double Wick rotation. In the next Section \ref{sec6}, we comment on the dual quantum mechanical model with broken scale invariance. Finally, in Section \ref{sec7}, we summarize our results with some future remarks.
%%%%%%%%%%%%%%%%%%%%%%%%%%%%%%%%%%%%%%%
\section{Krylov Operator Complexity: A brief introduction}
\label{sec:krylov_intro}

In quantum many-body systems, the operator growth serves as an important diagnostic tool for quantum chaos and information spreading. The Krylov operator complexity \cite{Parker:2018yvk} serves as the key element that quantifies this growth. In the following, we provide a brief introduction to the subject and discuss additional subtleties. 
%%%%%%%%%%%%%%%%%%%%%%%%%%%%%%%%%%%%%%%%%%%%%%%%%%%%
\subsection{Krylov Basis and Lanczos Algorithm}
We begin by considering an operator at $t=0$ i.e. $\mathcal{O}(0)$ in a quantum many body system that is governed by a Hamiltonian $H$. The time evolution of the operator in the Heisenberg picture is given by
\begin{align}
    \mathcal{O}(t) = e^{i H t} \mathcal{O}(0) e^{-i H t} = e^{i \mathcal{L} t} \mathcal{O}(0),
\end{align}
where $\mathcal{L} = [H, \, \cdot \,]$ stands for the Liouvillian superoperator. Taylor expanding $e^{i \mathcal{L} t} \mathcal{O}(0)$ yields the operator trajectory as a linear combination of nested commutators
\begin{align}
    \mathcal{O}(t) = \sum_{n=0}^{\infty} \frac{(it)^n}{n!} \mathcal{L}^n \mathcal{O}(0).
\end{align}
The subspace (of the full Hilbert space) spanned by these repeated applications of the Liouvillian,
\begin{align}
    \mathcal{K}_{\mathcal{O}} = \text{span} \left\{ \mathcal{O}(0), \mathcal{L} \mathcal{O}(0), \mathcal{L}^2 \mathcal{O}(0), \dots \right\},
\end{align}
is known as the Krylov subspace. 

The next step is to construct an orthonormal basis for $\mathcal{K}_{\mathcal{O}}$ following the Lanczos algorithm and using an appropriate inner product on the operator space (such as the Wightman or Frobenius inner product, $\langle A | B \rangle$). The algorithm proceeds iteratively as follows:
\begin{enumerate}
    \item Initialize the first Krylov basis state: 
    \begin{align}
        |K_0\rangle = \frac{|\mathcal{O}(0)\rangle}{\sqrt{\langle \mathcal{O}(0) | \mathcal{O}(0) \rangle}}, \quad b_0 \equiv 0.
    \end{align}
    \item For $n \ge 1$, compute the unnormalized state:
    \begin{align}
        |A_n\rangle = \mathcal{L} |K_{n-1}\rangle - b_{n-1} |K_{n-2}\rangle.
    \end{align}
    \item Determine the Lanczos coefficients $b_n$ and normalize to get the $n$-th Krylov basis element:
    \begin{align}
        b_n = \sqrt{\langle A_n | A_n \rangle}, \quad |K_n\rangle = \frac{|A_n\rangle}{b_n}.
    \end{align}
\end{enumerate}
In this Krylov basis $\{|K_n\rangle\}$, the Liouvillian $\mathcal{L}$ acts as a tridiagonal matrix, reducing the complex operator dynamics to a 1D semi-infinite tight-binding chain where the Lanczos cooefficients $b_n$ represent the hopping amplitudes between neighboring Krylov sites $\ket{K_n}$ and $\ket{K_{n+1}}$.

\subsection{Definition}
Krylov complexity is defined using the expansion of the time-evolved operator in the orthonormal Krylov basis \cite{Parker:2018yvk}
\begin{align}
    |\mathcal{O}(t)\rangle = \sum_{n=0}^{\infty} i^n \phi_n(t) |K_n\rangle.
\end{align}

Here, $\phi_n(t) = i^{-n} \langle K_n | \mathcal{O}(t) \rangle$ are the the real transition amplitudes, which represent the probability of finding the operator at site $n$ at time $t$, satisfying the normalization $\sum_n \phi_n(t)^2 = 1$. The wavefunctions $\phi_n(t)$ obey the discrete differential equation of the form
\begin{align}
    \dot{\phi}_n(t) = b_{n+1} \phi_{n+1}(t) - b_n \phi_{n-1}(t),
\end{align}
with an initial condition $\phi_n(0) = \delta_{n,0}$, which satisfies the normalization at $t=0$, that is, $\braket{\mathcal{O}(0)|\mathcal{O}(0)}=1$.

The Krylov complexity $\mathcal{C}(t)$ is defined as the average position of the wavepacket on the Krylov chain \cite{Parker:2018yvk}
\begin{align}
\label{10}
    \mathcal{C}(t) = \sum_{n=0}^{\infty} n \, |\phi_n(t)|^2.
\end{align}

\subsection{Universal Early-Time Growth}
At early times ($t \sim 0$), the wavepacket (or the state $\ket{\mathcal{O}(0)}$) remains localized near the origin $n=0$. Taylor expansion of these wavefunctions $\phi_n(t)$ follows as
\begin{align}
\label{11}
    \phi_0(t) &= 1 - \frac{1}{2} b_1^2 t^2 + \mathcal{O}(t^4), \\
    \phi_1(t) &= b_1 t + \mathcal{O}(t^3), \\
    \phi_n(t) &= \mathcal{O}(t^n) \quad \text{for } n \ge 2.
    \label{13}
\end{align}
Substituting these expansions \eqref{11}-\eqref{13} into \eqref{10}, gives the universal early-time quadratic growth
\begin{align}
    \mathcal{C}(t) = b_1^2 t^2 + \mathcal{O}(t^4).
\end{align}
Consequently, the rate of complexity growth exhibits a universal linear growth in the early-time regime
\begin{align}
    \partial_t \mathcal{C}(t) = 2 b_1^2 t + \mathcal{O}(t^3),
\end{align}
where the associated coefficient $b_1$ (also known as the first Lanczos coefficient) is directly related to the variance of the operator's energy spectrum.

\subsection{Subtleties in Continuum Quantum Field Theories}
\label{subsec:qft_subtleties}
We now discuss some additional subtleties in the context of continuum QFTs.
While extending the notion of Krylov complexity from discrete lattice models to continuum quantum field theories (QFTs), several non-trivial subtleties may emerge which we point out below.

\paragraph{UV Singularities and Regularization:} In continuum QFTs, the main subtlety is the short distance divergence while taking the inner product between local operators. This is due to coincident point singularities in two-point functions. To render the Lanczos coefficients $b_n$ finite and well-defined, one must therefore adopt a \textit{regularized} inner product. There are several proposals for this. The standard prescription is to introduce the Wightman inner product with thermal regularization,
    \begin{align}
    \label{16}
        \langle A | B \rangle_{\beta} = \text{Tr}\left( e^{-\frac{\beta H}{2}} A^\dagger e^{-\frac{\beta H}{2}} B \right) / Z(\beta).
    \end{align}
    
    The other proposal is to introduce operator smearing $O_\beta = e^{-\frac{\beta H}{4}} O e^{-\frac{\beta H}{4}}$, which dampens UV fluctuations.

    \paragraph{Continuous Spectrum and Unbounded Operator Growth:} The other issue is related to the large dimensionality of the Krylov subspace space in continuum field theories. Unlike finite-dimensional spin chains or matrix models where the Krylov space truncates at a finite dimension $K$, continuum QFTs feature an infinite-dimensional Hilbert space and a continuous Liouvillian spectrum. As a natural consequence of this, the sequence of Lanczos coefficients $b_n$ does not terminate.

    \paragraph{Dependence on the Choice of Inner Product:} Depending upon different choice of the inner products (e.g., Wightman, Kubo-Mori, or symmetric two-point functions), one might end up assigning distinct spectral weights to energy modes, which can alter the precise quantitative values of the Lanczos coefficients $b_n$, while preserving their asymptotic scaling behavior.

\subsection{Boundary States, Inner Product, and Motivation in the Ground-State Regime} 
Although we have introduced the inner product as in the previous discussion \eqref{16} at finite temperature, as a word of caution, it should be remembered that for the holographic backgrounds studied in this paper, this is not the case for the dual QFTs.
In contrast, our analysis is primarily focused with the zero-temperature (ground-state) sector of the dual QFT, as reflected by the dual gravitational backgrounds being deformed global $\text{AdS}$ spaces rather than thermal black hole geometries. In this particular context, the operator inner product corresponds to the vacuum expectation value 
\begin{align}
\langle A | B \rangle = \langle 0 | A^\dagger B | 0 \rangle,
\end{align}
regularized via suitable UV damping (or operator smearing) to prevent coincident-point divergences. Although chaotic signatures such as maximal exponential growth are not universal features of ground states, investigating Krylov complexity in this zero temperature deformed framework remains physically well motivated. In particular, our analysis isolates the pure \textit{geometric} effects of the integrable Yang-Baxter (YB) deformation on the kinematic operator growth and the effective dimensionality of the Krylov subspace, disentangling deformation induced chaos damping mechanisms from thermal scrambling effects inherent to black hole horizons.
%%%%%%%%%%%%%%%%%%%%%%%%%%%%%%%%%%%%%%%%%%%%%%%%%%%%%%%
\section{Deformed $ AdS_2 \times S^2 \times T^6 $ background in type IIB}
\label{sec3}
In the first part, we review the essential details about the background and its deformations. Next, we find perturbative solutions for the momentum ($P_\rho$) of the bulk particle, which is dual to an operator insertion in the UV field theory at the initial time $t \sim 0$. We plot the proper momentum ($P_\rho$) as a function of time that exhibits linear growth at early times.
%%%%%%%%%%%%%%%%%%%%%%%%%%%%%%%%%%%%%%%%%
\subsection{Embedding coordinates and the Krylov basis}
We begin by reviewing the deformed $ AdS_2 \times S^2 \times T^6 $ background in type IIB supergravity. These classes of geometries are constructed based on the notion of YB deformation \cite{Delduc:2013qra}-\cite{Delduc:2014kha} of the $AdS$ coset sigma model and is a generalization of the YB deformation of the Principal Chiral Model (PCM) \cite{Klimcik:2002zj}-\cite{Klimcik:2008eq}. The deformation is introduced by means of an $R$- matrix that satisfies the non-split modified YB- equation on the superisometry algebra $\mathfrak{psu}(1,1|2)$ of the undeformed background. It turns out that the resulting geometry, the dilaton, and the background fluxes solve a type IIB equation when all the simple roots of the Dynkin diagram are fermionic \cite{Hoare:2018ngg}. This in turn is related to the unimodularity of the $R$-matrix as suggested in \cite{Borsato:2016ose}. The resulting $R$-matrices (known as Drinfel’d-Jimbo R-matrices) correspond to a $q$-deformation of the super-isometry algebra \cite{Delduc:2016ihq}-\cite{Delduc:2013fga}.

The corresponding metric (in the string frame) can be expressed as \cite{Hoare:2018ngg}
\begin{align}
\label{e2.1}
&ds_{10}^2 = ds^2_4 + ds^2_{T^6}\\
&ds^2_4= \frac{L^2}{(1-\kappa^2 \varrho^2)}\Big(-(1+ \varrho^2)dt^2+\frac{d\varrho^2}{(1+ \varrho^2)} \Big)\nonumber\\
&+\frac{L^2}{(1+ \kappa^2 r^2)}\Big((1-r^2)d \phi^2 +\frac{dr^2}{(1-r^2)}  \Big)\\
&ds^2_{T^6} =L^2 d\varphi_i d\varphi_i ~;~ \kappa = \frac{2 \eta}{1- \eta^2}~;~i=4, \cdots , 9
\label{e2.3}
\end{align}
where $ \kappa \in [0,\infty]$ is the YB deformation parameter.

The dilaton associated with the background \eqref{e2.1}-\eqref{e2.3} is given by
\begin{align}
\label{e2.4}
e^{-2 \Phi}=e^{-2 \Phi_0}\frac{(1 - \kappa^2 \varrho^2)(1+ \kappa^2 r^2)}{1-\kappa^2(\varrho^2 - r^2 - \varrho^2 r^2)}.
\end{align}

To proceed further, we switch off the coordinates $\varphi_i \in T^6$ as they are not relevant for our analysis. Moreover, we restrict the geodesic of the massive particle in (deformed) $AdS_2$ where we set the angles on the two sphere as constants, namely $r = 0$ and $\phi=$ constant. This trivially fixes the dilaton $e^{-2 \Phi}=e^{-2 \Phi_0}=$ constant. In other words, the metric in the string frame is effectively the same as in the Einstein frame $ds^2_E =e^{-\frac{\Phi_0}{2}}ds^2_{string}$.

This finally yields the 2d metric in the Einstein frame as (after a rescaling by $e^{\frac{\Phi_0}{2}}$)
\begin{align}
\label{e2.5}
    d\tilde{s}^2_E= \frac{L^2}{(1-\kappa^2 \varrho^2)}\Big(-(1+ \varrho^2)dt^2+\frac{d\varrho^2}{(1+ \varrho^2)} \Big)
\end{align}
which serves as the starting point for our subsequent analysis. 

Notice that the background \ref{e2.5} is conformally equivalent to pure $AdS_2$ and smoothly transits into it in the limit $\kappa \rightarrow 0$. We now move on to the Krylov basis ($\ket{K_n}$) in a few steps. As a first step, we introduce $\tilde{\varrho}=\arctan \varrho$ and rewrite \ref{e2.5} as (we set $L=1$)
\begin{align}
\label{e2.6}
     d\tilde{s}^2_E=\Omega (\tilde{\varrho})(-dt^2 + d \tilde{\varrho}^2)~;~\Omega (\tilde{\varrho})=\frac{1}{\cos^2\tilde{\varrho}-\kappa^2 \sin^2 \tilde{\varrho}}.
\end{align}

In order to express the 2d metric \eqref{e2.6} in an appropriate radial coordinate ($\rho$), which is conjectured to be the dual of the Krylov basis $\Ket{K_n}$ associated with the 1d chain \cite{Caputa:2024sux}, we introduce the following change of variables:
\begin{align}
    \cos \tilde{\varrho}=\frac{1}{\cosh \rho}
\end{align}
which finally yields the 2d metric in the desired form
\begin{align}
\label{e2.8}
    d\tilde{s}^2_E=\Omega (\rho)(-\cosh^2 \rho dt^2 + d \rho^2)~;~\Omega(\rho)=\frac{1}{1-\kappa^2\sinh^2 \rho}.
\end{align}

The background \eqref{e2.8} is conformally equivalent to a ($0+1$)d CFT on a finite length $l=2\pi$. Notice that the conformal factor $\Omega (\rho)$ diverges at 
\begin{align}
\label{e2.9}
    \rho_B=\sinh^{-1}\Big(\frac{1}{\kappa}\Big)~;~\kappa >0
\end{align}
which defines the location of the holographic screen (or the effective boundary of the spacetime) \cite{Kameyama:2014via}-\cite{Kameyama:2014vma}. A closer inspection reveals that the location of the effective boundary \eqref{e2.9} shifts more towards the true boundary as $\kappa \sim 0$, which is the undeformed spacetime. 
%%%%%%%%%%%%%%%%%%%%%%%%%%%%%%%%%%%%%%%%%%%%%%%%%%%%%%
\subsection{Operator growth and Krylov complexity}
We perform analysis for the most generic value of the deformation parameter ($\kappa$). The particle is created near the screen ($\rho \sim \rho_B$) and as it probes deep inside the bulk, it corresponds to Krylov complexity growth at later times. The idea would be to compute the proper radial momentum ($P_\rho$) (which corresponds to the complexity growth \eqref{e1.1}) and plot it as a function of time ($t$) for different choices of the YB deformation parameter ($\kappa$).

The geodesic of the massive particle follows from the action
\begin{align}
    S=-m \int dt \sqrt{-g_{\mu \nu}\dot{x}^\mu \dot{x}^\nu}.
\end{align}

Using \eqref{e2.8}, one finds the following action for the massive particle
\begin{align}
    S=\int dt L ~;~L=-m \sqrt{\Omega(\rho)(\cosh^2\rho - \dot{\rho}^2)}.
\end{align}

The radial momentum of the particle can be expressed as
\begin{align}
\label{e2.12}
    P_{\rho}=\frac{m \dot{\rho} \sqrt{\Omega(\rho)}}{\sqrt{\cosh^2\rho - \dot{\rho}^2}}.
\end{align}

The Hamiltonian constraint, on the other hand, yields
\begin{align}
\label{e2.13}
    H=H_0= P_\rho \dot{\rho}-L=\frac{m \cosh^2\rho \sqrt{\Omega(\rho)}}{\sqrt{\cosh^2\rho - \dot{\rho}^2}}.
\end{align}

Using \eqref{e2.13}, we can further simplify \eqref{e2.12} as
\begin{align}
\label{e2.14}
    P_\rho =\frac{H_0 \dot{\rho}}{\cosh^2 \rho}.
\end{align}

Finally, we note down the equation of motion for $\rho(t)$, which turns out to be
\begin{align}
\label{e2.15}
    &\frac{d}{dt}\Big( \frac{\sqrt{\Omega}\dot{\rho}}{\sqrt{\cosh^2\rho -\dot{\rho}^2}}\Big)+\frac{\sqrt{\Omega}\sinh 2\rho}{2\sqrt{\cosh^2\rho -\dot{\rho}^2}}\nonumber\\
    &+\frac{\partial_t \Omega}{2 \dot{\rho}\sqrt{\Omega}}\sqrt{\cosh^2\rho -\dot{\rho}^2}=0.
\end{align}

A closer look reveals that \eqref{e2.15} is equivalent to setting the time derivative of the constraint \eqref{e2.13} equal to zero, that is, $\frac{d H_0}{dt}=0$. We therefore choose to work with the constraint condition \eqref{e2.13}, which after some simplification yields (with $m=1$)
\begin{align}
\label{e2.16}
    \dot{\rho}^2=\cosh^2 \rho \Big(1-\frac{\cosh^2\rho}{H^2_0}\Omega (\rho)  \Big).
\end{align}
%%%%%%%%%%%%%%%%%%%%%%%%%%%%%%%%%%%%%%%%%%%%
\subsubsection{The undeformed case}
Let us first revisit the undeformed case \cite{Caputa:2024sux} which sets $\Omega =1$. Taking into account that the particle starts with zero velocity near the boundary $\rho=\rho_\infty$ at initial time ($t=0$), we fix the constant of motion ($H_0$) in terms of the location of the asymptotic boundary ($\rho=\rho_\infty$)
\begin{align}
    H^2_0 = \cosh^2\rho_\infty
\end{align}
which is a large number in the true asymptotic limit $\rho_\infty \rightarrow \infty$.

This yields the following differential equation
\begin{align}
\label{e2.18}
    \dot{\rho}^2=\frac{\cosh^2 \rho}{\cosh^2\rho_\infty}(\cosh^2\rho_\infty - \cosh^2\rho).
\end{align}

Taking into account that $\rho_\infty \rightarrow \infty$, the above eq. \eqref{e2.18} yields the following solution
\begin{align}
\label{e2.19}
   \tan ^{-1}(\sinh \rho )=\frac{\pi}{2}-t +\mathcal{O}(H_0^{-2})
\end{align}
where the integration constant ($=\frac{\pi}{2}$) can be determined by setting $\rho =\infty$ at $t=0$. Notice that when $t=t_f=\frac{\pi}{2}$, the particle reaches the center ($\rho=0$) of $AdS_2$. 

One could rewrite the solution \eqref{e2.19} in the form of \cite{Caputa:2024sux} 
\begin{align}
\label{e2.20}
 \cos ^{-1}(\tanh \rho )=t \Rightarrow    \rho(t)=\rho_\infty \tanh ^{-1}\left(\cos t \right)+\mathcal{O}(H_0^{-2})
\end{align}
where the dual Krylov chain has a finite length $l=2\pi$ (also see eq. \eqref{e2.8}).

It turns out that, for the purpose of our subsequent discussion on YB effects, it would be important to estimate the above entity in eq. \eqref{e2.18} up to NLO
\begin{align}
    \tan ^{-1}(\sinh \rho )+\frac{1}{2H_0^2}\sinh\rho = \frac{\pi}{2}-t - \frac{C}{H^2_0}\equiv \hat{t}_f-t.
\end{align}

Clearly, with sub-leading corrections in action, the boundary is not strictly located at $\infty$ and therefore the particle takes a shorter time ($\hat{t}_f <t_f$) to reach the center of $AdS_2$. After some simple algebra, it is straightforward to obtain the solution
\begin{align}
    \sinh \rho = \alpha \cot \Big( t+ \frac{C}{H^2_0}\Big)~;~\alpha=1-\frac{1}{2 H^2_0}
\end{align}
which is the generalization of \eqref{e2.19} in the presence of a finite radial cut-off. 
%%%%%%%%%%%%%%%%%%%%%%%%%%%%%%%%%%%%%%%%%%%%
\subsubsection{The YB effects}
In the context of pure $AdS_2$, the above analysis does not make much sense since the constant of motion $H_0=\infty$ and the boundary is strictly located at asymptotic infinity. However, in the presence of the YB deformation, the boundary (or the holographic screen) is located at a finite radial cut-off \eqref{e2.9} depending on the value of the deformation parameter ($\kappa$).

Like before, by considering the fact that the particle is created near the holographic screen at $\rho \sim \rho_\Lambda \sim \rho_\infty$, we find the velocity of the particle for small deformation ($\kappa \ll 1$)
\begin{align}
\label{e2.23}
    \dot{\rho}=\kappa\cosh\rho \sinh \rho|_{\rho \sim \rho_\Lambda} +\mathcal{O}(\kappa^2)
\end{align}
where $\rho_\Lambda <\rho_B$. In other words, the effect of YB deformation ($\kappa \neq 0$), is to set a non-zero (small but finite) value for the radial velocity ($\dot{\rho}<1$) and hence the proper momentum ($P_\rho$), which brings a crucial difference with the analysis in the undeformed case. 

In order to explore the effects of the YB deformation on the Krylov complexity, one has to take into account the terms at $\mathcal{O}(H^{-2}_0)$, which yields an integral equation of the form
\begin{align}
\label{e2.24}
    \int \frac{d\rho}{\cosh \rho}+\frac{1}{2 \kappa H_0^2}\int d\rho \frac{\cosh\rho}{1- \kappa^2 \sinh^2\rho}+S_{ct}=\hat{t}_f-t
\end{align}
where $\hat{t}_f$ is the time that the particle takes to reach the center of $AdS_2$.

Notice that the second integral diverges near the boundary $\rho =\rho_B$. On the other hand, in the limit $\kappa \rightarrow 0$, one precisely recovers the undeformed eq. \eqref{e2.19}. The YB effect therefore amounts to a regularization and the addition of a counter term. After performing a careful analysis of the integral, one finds the counter term of the following form 
\begin{align}
\label{e2.25}
    &S_{ct}=-\frac{a}{2 \kappa H_0^2}\lim_{\rho \rightarrow \rho_B}\int_{\rho \sim \rho_B} d\rho \frac{\cosh\rho}{1- \kappa^2 \sinh^2\rho}\nonumber\\
    &=\frac{a}{4 \kappa^2 H_0^2}\int_{z\sim \epsilon}\frac{dz}{z} =\frac{a}{4 \kappa^2 H_0^2}\log \epsilon
\end{align}
where $a$ is the appropriate pre-factor that we fix below in order to cancel the divergence. Here, in the above derivation, we use the fact that near the holographic screen $1- \kappa^2 \sinh^2\rho |_{\rho \sim \rho_B}=2 (1-\kappa \sinh \rho)|_{\rho\sim \rho_B}$, where $\rho_B$ is defined in \eqref{e2.9}. 

Using \eqref{e2.25} and performing the integral in eq. \eqref{e2.24}, we obtain the following relation
\begin{align}
\label{e2.26}
    \tan ^{-1}(\sinh \rho )+\frac{1}{2 \kappa H_0^2}\tanh ^{-1}(\kappa  \sinh \rho )+S_{ct}=\hat{t}_f-t.
\end{align}

After a careful analysis, it can be shown that the time for the massive particle to reach the center of $AdS_2$ turns out to be\footnote{In what follows, in the subsequent analysis, we always absorb the counter term \eqref{e2.25} in the definition of $\hat{t}_f$, meaning that the resulting entity is always finite.}  
\begin{align}
\label{e2.27}
    &\hat{t}_f = \tan ^{-1}(1/\kappa )+\frac{1}{2 \kappa H_0^2}\tanh ^{-1}(1-z)\Big|_{z \sim \epsilon}+S_{ct}\nonumber\\&=\tan ^{-1}(1/\kappa )+\frac{\log 2}{4 \kappa H_0^2}
\end{align}
where the second term is purely sourced due to YB correction in bulk supergravity. Notice that in order to cancel the divergence in eq. \eqref{e2.27}, one has to set $a=\kappa$. Finally, it should be noted that in the limit $\kappa \rightarrow 0$, one recovers the undeformed theory \eqref{e2.19}. This can be seen by taking the limit $\kappa \rightarrow 0$ in the second term in \eqref{e2.27}, where we denote the Hamiltonian constraint near the boundary as $H^2_0 = \cosh^2\rho \Omega (\rho)\Big|_{\rho\sim \rho_B}=(1+\kappa^{-2})\Omega (\rho)\Big|_{\rho\sim \rho_B}$.
%%%%%%%%%%%%%%%%%%%%%%%%%%%%%%%%%%%%%%%%%%%%%%%%%%%%%%
\paragraph{Small deformation limit.} An exact analytical solution of eq. \eqref{e2.26} is a nontrivial task. However, a solution can be obtained in the limit where the YB effects are small enough to consider a perturbative expansion in the parameter $\kappa(\ll 1)$. This yields the following 
\begin{align}
    \tan ^{-1}(\sinh \rho )+\frac{\kappa ^2}{2}  \sinh \rho =\hat{t}_f-t
\end{align}
where the time to reach the bulk interior turns out to be
\begin{align}
\label{e2.29}
    &\hat{t}_f=\frac{\pi }{2}-\left| \kappa \right|+\frac{\kappa }{4}\log 2+\mathcal{O}(\kappa^3)= \frac{\pi }{2}-\beta \nonumber\\&\beta=\left| \kappa \right|-\frac{\kappa }{4}\log 2+\mathcal{O}(\kappa^3).
\end{align}
Notice that the time \eqref{e2.29} to reach the center of the bulk is shorter compared to the undeformed case \eqref{e2.19}. This comes from the fact that the boundary is now shifted towards the interior and is at a finite radial cut-off depending on the value of the YB parameter ($\kappa$).

Finally, we have a solution for the radial coordinate
\begin{align}
\label{e2.30}
    &\rho(t) = \tanh^{-1}\Big( \frac{\gamma \cot \bar{t}}{\sqrt{1+\gamma^2 \cot^2\bar{t}}} \Big)\\
    &\bar{t}=t+\beta ~;~ \gamma = (1+\frac{\kappa^2}{2})^{-1}
    \label{48}
\end{align}
that smoothly transits into the undeformed solution \eqref{e2.20}, in the limit $\kappa \rightarrow 0$.

Using \eqref{e2.30}, we finally express the proper momentum \eqref{e2.14} as\footnote{Here, $H_0$ keeps the information about the UV scale of the dual theory \cite{Caputa:2024sux}.}
\begin{align}
\label{e2.32}
    P_\rho /H_0 = -\frac{\gamma  \csc ^2(\beta +t)}{\left(\gamma ^2 \cot ^2(\beta +t)+1\right)^{3/2}}.
\end{align}

Setting $\gamma =1$ and $\beta =0$ in \eqref{e2.32}, one recovers the results of \cite{Caputa:2024sux}
\begin{align}
    P_\rho/H_0 =-\sin t
\end{align}
in units where $m=1$ and $l=2\pi$, which we have set at the beginning of our computation. 

\begin{figure}[htbp]
    \centering
    \includegraphics[width=0.5\textwidth]{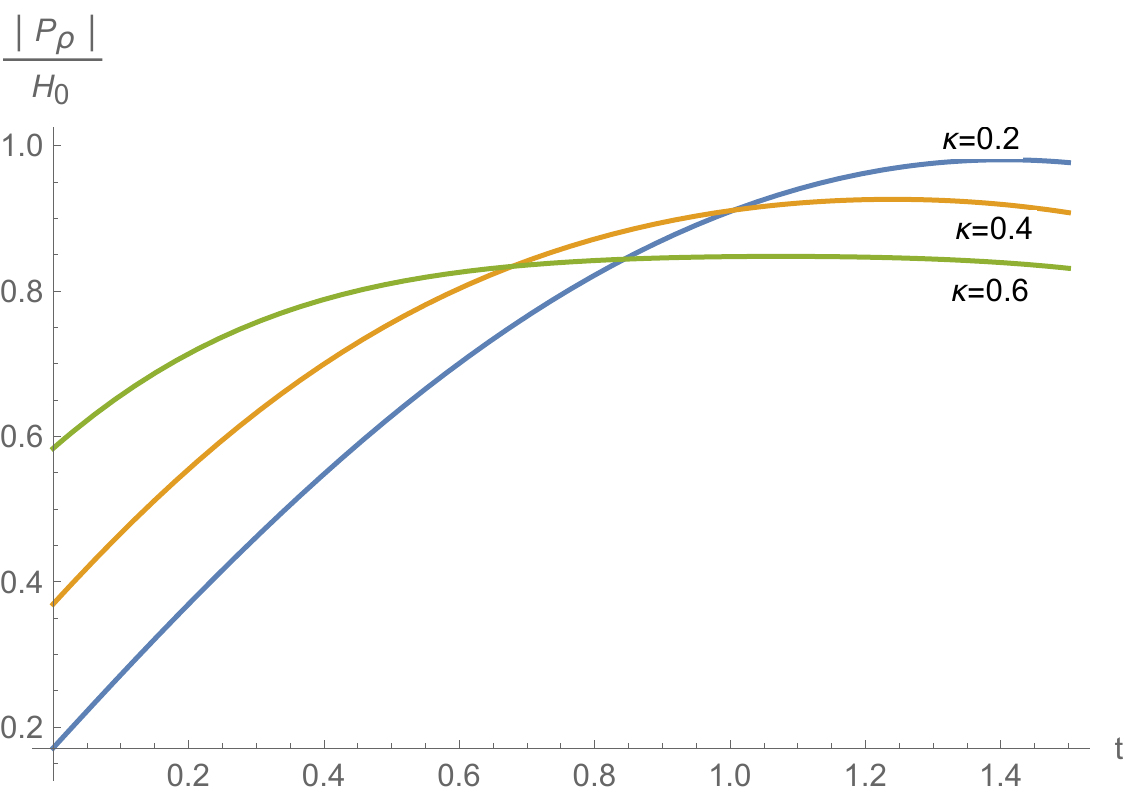}
    \caption{We show the rate of growth of Krylov complexity for various choices of the YB deformation parameter ($\kappa$).}
    \label{fig1}
\end{figure}

%\begin{figure}
    %\centering
    %\includegraphics[width=0.5\textewidth]{fig1.pdf}
    %\caption{We show the rate of growth of Krylov complexity for various choices of the YB deformation parameter ($\kappa$). }
    %\label{fig1}
%\end{figure}

A closer observation reveals that the effect of the YB deformation is to reduce the rate of growth of Krylov complexity, which is also shown in Fig.\ref{fig1}. This can be understood as a fact that the span of the radial direction ($\rho$) becomes smaller with a larger deformation, which reduces the effective dimension of the Hilbert space associated with the Krylov basis. 

Finally, as a second important observation and as also discussed below eq.\eqref{e2.23}, we notice that the particle must have a non-zero proper momentum (near the location of the holographic screen) in order to see the effects due to YB deformation, which is shown in Fig.\ref{fig1} and can be seen from the following expansion of \eqref{e2.30}
\begin{align}
\label{51}
    &-P_\rho/H_0 =\sin t -\frac{\kappa}{4}\cos t (-4+\log 2)\nonumber\\&+\frac{\kappa ^2 }{32 }\sin t \left(24 \cos (2 t)-8-\log ^22+\log (256)\right)+\mathcal{O}(\kappa^3)
\end{align}
which is non-zero at the initial time $t=0$ and is proportional to $\kappa$.

One way to interpret the above result is to argue that the YB deformation captures a subspace of the full Hilbert space associated with Krylov complexity growth. The time at which one starts measuring the complexity growth is not the true initial time. From the perspective of the undeformed theory on the dual gravity side, this can be understood by noting the fact that the YB deformation forces the particle to be detected at a finite radial distance from the true asymptotic boundary at infinity. As a result, the particle is found with a finite momentum that corresponds to a non-zero complexity growth to begin with. 
 \paragraph*{Modified momentum/complexity correspondence:}
There is a nicer way to interpret the momentum/complexity correspondence for YB deformed geometries, which could be thought of as the ``YB modified'' version of the original proposal \cite{Caputa:2024sux}. Expanding the R.H.S. of \eqref{51} for small times ($t \sim 0$) and retaining ourselves up to leading order in $\kappa \sim 0$, it is straightforward to show
\begin{align}
\label{52}
   -P_\rho/H_0|_{\bar{t} \sim 0}=t+\kappa - \frac{\kappa}{4}\log 2+\mathcal{O}(\kappa^2) = \bar{t}+\mathcal{O}(\kappa^2).
\end{align}
In other words, in this modified version of the correspondence, the proper momentum \eqref{52} indeed varies linearly with respect to the ``time translated'' parameter \eqref{48}, where the amount of translation $\beta$ \eqref{e2.29} is a measure of the YB effect. This naturally leads to the notion of complexity that goes quadratically with the rescaled time, that is, $C(\bar{t})\sim \bar{t}^2$, in the limit when $\kappa \ll 1$.
%%%%%%%%%%%%%%%%%%%%%%%%%%%%%%%%%%%%%%%%%%%%%%%%%%%%%%%
\section{Two parameter deformation of $ AdS_3 \times S^3 \times T^4 $}
\label{sec4}
These backgrounds are in spirit similar to those discussed in the previous section, in the sense that they are obtained as a result of integrable deformation of type II Green-Schwarz (GS) superstrings by means of $R$-matrices satisfying non-split modified classical YB equation on the super-isometry algebra of the undeformed background. Like in the previous example, these $R$-matrices satisfy the unimodularity condition that ensures a supergravity solution as a target space geometry of the string. Below, we review different supergravity solutions \cite{Seibold:2019dvf} and thereby proceed towards constructing a possible Krylov basis for it.
%%%%%%%%%%%%%%%%%%%%%%%%%%%%%%%%%%%%%%%%%%%%%
\subsection{Supergravity solutions and Krylov basis}
To begin with, $ AdS_3 \times S^3 \times T^4 $ superstrings allow a superisometry algebra to be in the form of a product $\hat{G}_L\times \hat{G}_R$, which allows for a two parameter deformation \cite{Hoare:2014oua}, much similar in spirit as that of the bi-Yang-Baxter deformation of the sigma model \cite{Klimcik:2014bta}. Corresponding to each copy of the superisometry algebra, one can introduce a real deformation parameter $\kappa_L$ or $\kappa_R$, which results in a two parameter family of integrable deformations of type II GS superstrings and an associated supergravity solution. Each family of supergravity solution corresponds to a particular unimodular $R$-matrix associated with fully fermionic Dynkin diagram. Interestingly, one of these backgrounds exhibits mirror symmetry \cite{Arutyunov:2014cra}. We explore the complexity growth for the mirror background towards the end of this paper.

The superisometry algebra we are interested in is constructed using one copy of the supergroup $\hat{G}=PSU(1,1|2)$, which corresponds to strings moving in $ AdS_3 \times S^3 \times T^4 $ in the presence of background fluxes. There happen to be two inequivalent classes of unimodular $R$-matrices, giving rise to two distinct classes of supergravity solutions that have a unique metric and different dilaton profiles \cite{Seibold:2019dvf}. For our purpose, we ignore the background RR and the NS fluxes. These backgrounds can be classified as follows:
\paragraph{Class I background.} The metric and the dilaton are given by
\begin{align}
\label{e3.1}
    ds^2_{10}&=\frac{1}{F(\varrho)}\Big[ -(1+\varrho^2)(1+\kappa^2_-(1+\varrho^2))dt^2\nonumber\\
    &+\frac{d \varrho^2}{(1+\varrho^2)}+\varrho^2(1- \kappa^2_+ \varrho^2)d\psi^2 \nonumber\\& +2 \kappa_- \kappa_+ \varrho^2 (1+\varrho^2)dt d\psi\Big]\nonumber\\
    &+\frac{1}{\tilde{F}(r)}\Big[(1-r^2)(1+\kappa^2_-(1-r^2))d \varphi^2+\frac{dr^2}{1-r^2}\nonumber\\&+r^2(1+ \kappa^2_+ r^2)d\phi^2+2 \kappa_- \kappa_+ r^2 (1-r^2)d\varphi d\phi \Big]+dx^i dx^i\\
    &F(\varrho)=1+\kappa^2_-(1+\varrho^2)-\kappa^2_+ \varrho^2\nonumber\\
    &\tilde{F}(r)=1+\kappa^2_-(1-r^2)+\kappa^2_+ r^2\\
    &e^{-2\Phi}=e^{-2\Phi_0}\frac{F(\varrho)\tilde{F}(r)}{P(\varrho ,r)^2}\nonumber\\
    &P(\varrho ,r)=1-\kappa^2_+(\varrho^2 -r^2 -\varrho^2 r^2)+\kappa^2_-(1+\varrho^2)(1-r^2)
    \label{e3.3}
\end{align}
where we denote the deformation parameters as $\kappa_{\pm}=\frac{1}{2}(\kappa_L \pm \kappa_R)$.

The dilaton depends on the choice of the unimodular $R=\text{diag} (R_L, R_R)$-matrix, where $R_L$ and $R_R$ individually satisfy the modified classical YB equation on the undeformed $\mathfrak{psu}(1,1|2)_L$ and $\mathfrak{psu}(1,1|2)_R$ of the respective superisometry groups. For the background dilaton \eqref{e3.3} above, we have the choice of the $R$- matrix as $R_1=\text{diag}(R_{\mathbb{P}_1},-R_{\mathbb{P}_1})$, where $R_{\mathbb{P}_i}$ are the new $R$-matrices obtained using the permutation operators $\mathbb{P}_i$ on the reference $R$- matrix $R_0$ \cite{Seibold:2019dvf}. Notice that each $R_{\mathbb{P}_i}$ is associated with the full fermionic Dynkin diagram, and the matrix $R_1$ is defined on the superisometry algebra $\mathfrak{psu}(1,1|2)_L\oplus \mathfrak{psu}(1,1|2)_R$.

For our purpose, we choose a particular geodesic, where we set $x^i=0$, $r=0$, $\varphi =\phi=$ constants. The corresponding metric that the particle feels (in the string frame) is given by
\begin{align}
\label{e3.4}
    &ds_3^2=\frac{1}{F(\varrho)}\Big[ -(1+\varrho^2)(1+\kappa^2_-(1+\varrho^2))dt^2\nonumber\\
    &+\frac{d \varrho^2}{(1+\varrho^2)}+\varrho^2(1- \kappa^2_+ \varrho^2)d\psi^2 \nonumber\\& +2 \kappa_- \kappa_+ \varrho^2 (1+\varrho^2)dt d\psi\Big]\\
    &F(\varrho)=1+\kappa^2_-(1+\varrho^2)-\kappa^2_+ \varrho^2~;~\tilde{F}(r)=1+\kappa^2_-\\
    &e^{-2 \Phi}=e^{-2 \Phi_0}\frac{(1+\kappa^2_-)}{F(\varrho)}.
\end{align}

After appropriate rescaling, the metric in the Einstein's frame reads as
\begin{align}
\label{e3.7}
     d\tilde{s}^2_E &=\frac{1}{F^{5/4}(\varrho)}\Big[ -(1+\varrho^2)(1+\kappa^2_-(1+\varrho^2))dt^2+\frac{d \varrho^2}{(1+\varrho^2)}\nonumber\\
     &+\varrho^2(1- \kappa^2_+ \varrho^2)d\psi^2 \nonumber\\& +2 \kappa_- \kappa_+ \varrho^2 (1+\varrho^2)dt d\psi\Big].
\end{align}

Finally, by introducing a change of variable, 
\begin{align}
\label{e3.8}
    1+\varrho^2=\cosh^2\rho
\end{align}
one can re-express the above metric \eqref{e3.7} as
\begin{align}
    &d\tilde{s}^2_E =\frac{1}{F^{5/4}(\rho)}\Big[ ds^2_{AdS_3}+ds^2_{deform}\Big]\\
    \label{e3.10}
    &ds^2_{AdS_3}=-\cosh^2\rho dt^2 +d\rho^2 +\sinh^2\rho d\psi^2\\
    \label{e3.11}
    &ds^2_{deform}=-(-\kappa_- \cosh^2 \rho dt+\kappa_+ \sinh^2\rho d\psi)^2\\
    &F(\rho)=1+\kappa^2_- \cosh^2\rho -\kappa^2_+ \sinh^2\rho.
    \label{e3.12}
\end{align}

Clearly, \eqref{e3.10} is the usual piece that corresponds to a 2d CFT (on a finite length $l=2\pi$) associated with the dual Krylov basis \cite{Caputa:2024sux}. This basis is deformed by the second term \eqref{e3.11} in the bulk, whose dual interpretation is not very clear at the moment. The boundary of the spacetime is given by the vanishing of the scale factor
\begin{align}
\label{e3.13}
    F(\rho =\rho_B)=0 \Rightarrow \rho_B=\sinh^{-1}\Big(\frac{\sqrt{1+\kappa^2_-}}{\sqrt{\kappa^2_+ -\kappa^2_-}} \Big).
\end{align}

\paragraph{Class II background.} The second class background corresponds to a different choice of the unimodular $R_2=\text{diag}(R_{\mathbb{P}_1},-R_{\mathbb{P}_2})$ matrix \cite{Seibold:2019dvf} which corresponds to the background metric as in \eqref{e3.1}. The difference appears in the background dilaton, which is given by
\begin{align}
\label{e3.14}
    &e^{-2\Phi}=e^{-2\Phi_0}\frac{F(\varrho)\tilde{F}(r)}{P(\varrho ,r)^2}\nonumber\\
    &P(\varrho ,r)=1-\kappa^2_+\varrho^2 r^2 +\kappa^2_-(1+\varrho^2 r^2).
\end{align}

Setting $r=0$ and following the steps mentioned above in eq.\eqref{e3.4}, we obtain the metric of the subspace in Einstein's frame (after a suitable rescaling) as
\begin{align}
\label{e3.15}
    &d\tilde{s}^2_E =\frac{1}{F^{3/4}(\rho)}\Big[ ds^2_{AdS_3}+ds^2_{deform}\Big]
\end{align}
where the entities in parentheses are defined in \eqref{e3.10}-\eqref{e3.11}. Notice that the background \eqref{e3.15} differs from the previous solution only in the scale factor ($F(\rho)$), which now appears with a different power. However, qualitatively these two solutions should give rise to identical physics when studying the Krylov complexity in the dual CFT. 
%%%%%%%%%%%%%%%%%%%%%%%%%%%%%%%%%%%%%%%%%%%%%
\subsection{Operator growth and Krylov complexity}
Following \cite{Caputa:2024sux}, we set $\psi=$ constant, which yields the massive particle action 
\begin{align}
\label{e3.16}
    &S= \int dt L~;~L=-\frac{m}{F^{n}(\rho)}\sqrt{K(\rho)\cosh^2\rho -\dot{\rho}^2}\\
    &K(\rho)=1+\kappa^2_- \cosh^2\rho.
\end{align}

Notice that the function $K(\rho)$ sources the deformation of the dual $SL(2,R)$ Krylov chain on the boundary. In the special case, where the deformation parameters associated with both left and right $PSU(1,1|2)$ are identical ($\kappa_L=\kappa_R$), that is, $\kappa_-=0$, we have a situation encountered in the previous section. In other words, this would refer to a background that is conformally equivalent to an $AdS_2$ subspace that is dual to the $SL(2,R)$ Krylov chain. Finally, here $n$ is a free parameter that classifies different supergravity backgrounds, namely $n=\frac{5}{8}$ for the class I background and $n=\frac{3}{8}$ for the class II background.

Next, we note down the radial momentum of the particle
\begin{align}
\label{e3.18}
    P_\rho = \frac{m \dot{\rho}}{F^n(\rho)\sqrt{K(\rho)\cosh^2\rho -\dot{\rho}^2}}.
\end{align}

The Hamiltonian of the particle, on the other hand, turns out to be
\begin{align}
\label{e3.19}
    H=H_0=P_\rho \dot{\rho}-L=\frac{m}{F^n (\rho)}\frac{K(\rho)\cosh^2\rho}{\sqrt{K(\rho)\cosh^2\rho -\dot{\rho}^2}}.
\end{align}

Using \eqref{e3.19}, one can simplify the proper momentum \eqref{e3.18}
\begin{align}
\label{e3.20}
    P_\rho = \frac{H_0 \dot{\rho}}{K(\rho)\cosh^2\rho}.
\end{align}

Finally, we note down the equation of motion for $\rho(t)$ that follows from \eqref{e3.16}
\begin{align}
\label{e3.21}
    &\frac{d}{dt}\Big( \frac{1}{F^n (\rho)}\frac{\dot{\rho}}{\sqrt{K(\rho)\cosh^2\rho -\dot{\rho}^2}}\Big)\nonumber\\
    &-\frac{n \partial_\rho F}{F^{n+1}(\rho)}\sqrt{K(\rho)\cosh^2\rho -\dot{\rho}^2}\nonumber\\
    &+\frac{(\partial_\rho K \cosh^2\rho +K(\rho)\sinh 2\rho)}{2F^n(\rho)\sqrt{K(\rho)\cosh^2\rho -\dot{\rho}^2}}=0.
\end{align}

Using the constraint \eqref{e3.19}, we can further simplify the above eq.\eqref{e3.21}
\begin{align}
\label{e3.22}
     &\frac{d}{dt}\Big( \frac{\dot{\rho}}{K(\rho)\cosh^2\rho}\Big)-\frac{n \partial_\rho F}{\bar{H}^2_0}\frac{K(\rho)\cosh^2\rho}{F^{2n+1}(\rho)}\nonumber\\
     &+\frac{1}{2}\partial_\rho \log K(\rho)+\tanh\rho=0
\end{align}
where $\bar{H}_0=\frac{H_0}{m}$ which can be read from eq.\eqref{e3.19}. 

Taking into account the fact that near the boundary\footnote{Here, the cut-off is set such that $\kappa_{\pm}\cosh \rho_\Lambda < 1$ in the limit where $\rho_\Lambda \gg 1$ and $\kappa_{\pm}\ll 1$.} $\rho_\Lambda < \rho_B$ and in the limit of small deformation, that is, $\kappa^2_-\sim 0$, together with the fact $\dot{\rho}(t=0)\ll 1$, the Hamiltonian $\bar{H}_0 \sim \cosh \rho_\Lambda \gg 1$ is a large number. In other words, it is legitimate to drop the second term in \eqref{e3.22} and solve the rest of the equation, which can be expressed as
\begin{align}
\label{e3.23}
    \frac{d^2}{dt^2}\tanh \rho -\frac{d}{dt}\tanh \rho\frac{d}{dt}\log K +\frac{1}{2}\partial_\rho K+K\tanh\rho=0.
\end{align}

Two points should be noted here. First, the solution of \eqref{e3.23} is the same for both classes I and II backgrounds since their information is hidden in the function $F(\rho)$ which appears in the NLO in the expansion $1/\bar{H}^2_0$. Second, the solution depends on one of the parameters $\kappa_-$ that appears through the function $K(\rho)$. In other words, in the limit $\kappa_-=0$, we have a situation identical to that of \cite{Caputa:2024sux}, which yields a solution of the form\footnote{Notice that, setting $\kappa_-=0=\frac{1}{2}(\kappa_L -\kappa_R)$ does not ensure that the deformations $\kappa_{L,R}$ are individually zero; this could be one of the possibilities, but the other possibility is that they are equal, namely $\kappa_L=\kappa_R$ \cite{Seibold:2019dvf}. At LO in $1/\bar{H}^2_0$ expansions, these two situations cannot be differentiated at the level of the equations of motion. At NLO, these effects can be differentiated in the presence of the parameter $\kappa_+=\frac{1}{2}(\kappa_L +\kappa_R)$. However, a non-zero $\kappa_-$ certainly corresponds to non-vanishing YB deformation which we explore here.}
\begin{align}
    \tanh\bar{\rho}(t)=\tanh \rho_\Lambda \cos t.
\end{align}

In the presence of the deformation and considering it to be small enough\footnote{This would correspond to setting the left and the right deformation parameters close to each other, that is, $\kappa_L \sim \kappa_R$ while both are non-zero, $\kappa_L \neq \kappa_R \neq 0$.} $\kappa^2_- \ll 1$, we can find an approximate solution of \eqref{e3.23}. We propose a solution of the form
\begin{align}
    \tanh \rho = \tanh \bar{\rho}+\kappa^2_- y(t)+\mathcal{O}(\kappa^4_-).
\end{align}

The equation at $\mathcal{O}(\kappa^2_-)$ takes the following form
\begin{align}
    \frac{d^2y}{dt^2}+y=0~;~\Rightarrow y(t)=A \cos t + B \sin t.
\end{align}

To fix these constants, we notice that in the presence of YB deformations, $\rho(t=0)=\rho_\Lambda$, which yields $A=0$. Therefore, one is left with the remaining constant $B=\frac{\dot{\rho}(0)}{\kappa^2_-\cosh^2\rho_\Lambda}$, which is fixed in terms of the non-zero velocity at the boundary. 

Therefore, the complete solution at NLO reads
\begin{align}
\label{e3.27}
    \tanh\rho =\tanh \rho_\Lambda \cos t+\kappa^2_- B \sin t.
\end{align}
\begin{figure}[htbp]
    \centering
    \includegraphics[width=0.5\textwidth]{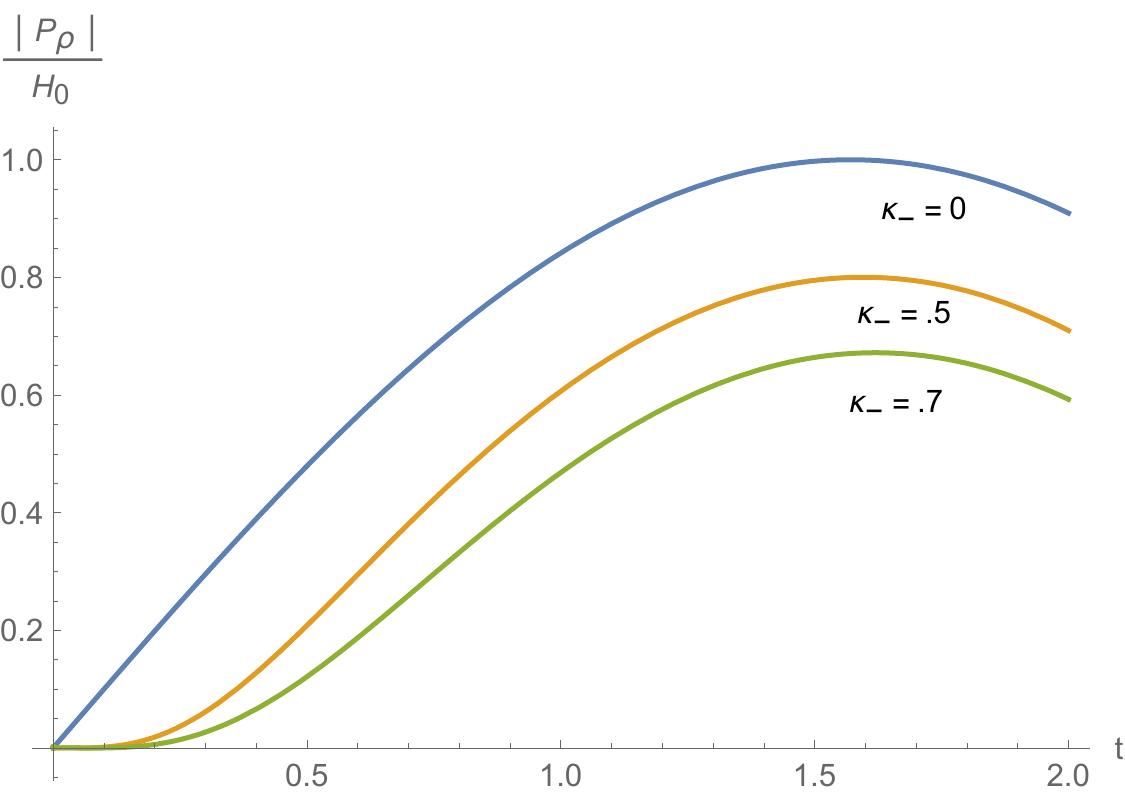}
    \caption{We plot proper momentum against time for different choices of the deformation parameter $\kappa_-=\frac{1}{2}(\kappa_L-\kappa_R)$, where $\kappa_L$ and $\kappa_R$ are close to each other. When they are exactly equal, we have the massive particle experiences an emerging $SL(2,R)$ symmetry which is therefore identical to that of the original notion of the Krylov complexity in a Krylov basis. In our analysis, we set $B=0.1$ (which accounts for the initial velocity ($\dot{\rho}(0)$) of the particle, which is very small to begin with) and $\rho_\Lambda =10$. For $\kappa_- \neq 0$, the initial momenta are very small ($\sim \mathcal{O}(10^{-10})$) and cannot be distinguished from the momenta with $\kappa_-=0$, which is actually zero namely $P_\rho(t=0)|_{\kappa_-=0}=0$.}
    \label{figads3}
\end{figure}

Using \eqref{e3.27}, we finally express the proper momentum \eqref{e3.20} as
\begin{align}
    &P_\rho/H_0=\frac{ (\kappa_- ^2 B \cos t-\tanh \rho_\Lambda \sin t)}{(1+\kappa^2_- D(t))}\\
    &D^{-1}=1-\left(\kappa_- ^2 B \sin t +\tanh \rho_\Lambda \cos t\right)^2.
\end{align}

Considering an expansion in the deformation parameter $\kappa^2_- \sim \kappa_L - \kappa_R$, we obtain
\begin{align}
\label{e3.30}
    &P_\rho/H_0=-\tanh \rho_\Lambda \sin t \nonumber\\
    &+\kappa_- ^2 \left(B \cos t-\frac{\tanh \rho_\Lambda \sin t}{\tanh ^2\rho_\Lambda \cos ^2 t-1}\right)+\mathcal{O}(\kappa_- ^4)
\end{align}
where the leading term in \eqref{e3.30} mimics the result of the undeformed theory \cite{Caputa:2024sux}. Clearly, setting $t=0$, one finds  a non-zero momentum, namely $P_\rho/H_0 \sim \kappa^2_- B \ll 1 $ in the limit $\kappa_L \sim \kappa_R$. This can also be seen from Fig.\ref{figads3}, where we plot the proper momentum for different choices of the deformation parameter ($\kappa_-$). Notice that the case $\kappa_-=0$ is identical to that of the Krylov complexity \cite{Caputa:2024sux}. As deformation ($\kappa_-$) increases, the rate of growth is suppressed. In the plot, it is difficult to discriminate between the initial momenta as they are vanishingly small in the range of parameters ($\rho_\Lambda\gg 1$ and $B<1$) chosen for the plot.
%%%%%%%%%%%%%%%%%%%%%%%%%%%%%%%%%%%%%%%%%%%%%%%%%
\section{Comments on the mirror model}
\label{sec5}
The mirror dual \cite{Arutyunov:2007tc}-\cite{Arutyunov:2014cra} of the class II background has been constructed in \cite{Seibold:2019dvf}. Setting $\kappa_+=\kappa$ and $\kappa_-=0$, one finds the metric and the background dilaton as
\begin{align}
\label{e4.1}
    &ds^2=-f_1(r)dt^2+f_2(\varrho)d\varrho^2+\varrho^2 d\psi^2 \nonumber\\&+f_3 (\varrho)d\varphi^2+f_4(r)dr^2+r^2 d\phi^2+(dx^i)^2\\
    &f_1(r)=\frac{(1+ \kappa^2 r^2)}{(1-r^2)}~;~f_2(\varrho)=\frac{(1+\varrho^2)^{-1}}{(1- \kappa^2 \varrho^2)}\nonumber\\&f_3(\varrho)=\frac{(1-\kappa^2 \varrho^2)}{(1+\varrho^2)}~;~f_4(r)=\frac{(1-r^2)^{-1}}{(1+\kappa^2 r^2)}\\
    &e^{-2 \Phi}=e^{-2 \Phi_0}\frac{(1-r^2)(1+\varrho^2)}{(1-\kappa^2 \varrho^2 r^2)^2}.
    \label{e4.3}
\end{align}

Comparing, for example, the $tt$ component of the metric in \eqref{e3.1} and in \eqref{e4.1}, one finds that they can be identified (with $\kappa_-=0$) under transformation $\varrho \rightarrow \varrho/\kappa$, $\tilde{\kappa}=1/\kappa$ and thereby identifying $\varrho \leftrightarrow r$. On a similar note, the background dilaton in \eqref{e3.14} can be mapped to \eqref{e4.3}, which is an artifact of the mirror duality.

For the geodesic under consideration, we set $\phi=\varphi=x^i=0$, which reveals the metric in the string frame as
\begin{align}
    ds^2_{string}=-f_1(r)dt^2+f_2(\varrho)d\varrho^2+\varrho^2 d\psi^2+f_4(r)dr^2.
\end{align}

For the point particle geodesic, one has to work with the metric in Einstein's frame
\begin{align}
\label{e4.5}
    d\tilde{s}^2_E=e^{-\frac{\Phi}{2}}ds^2_{string}.
\end{align}
%%%%%%%%%%%%%%%%%%%%%%%%%%%%%%%%%%%%%%%%%%%%%
\subsection{Krylov basis and geodesic motion}
The Krylov basis can be obtained following a double Wick rotation $t \rightarrow -i \psi$ and $\psi \rightarrow i t$ and thereby considering an expansion close to $r\sim 0$. The above operation corresponds to a rotation of the cylinder by $\frac{\pi}{2}$, which converts any timelike interval to a spacelike interval. The Euclidean time has periodicity $\beta=2\pi$, which therefore corresponds to a CFT$_2$ at a finite temperature \cite{Caputa:2024sux}. Using the map \eqref{e3.8}, the corresponding metric \eqref{e4.5} is given by
\begin{align}
\label{e4.6}
    &d\tilde{s}^2_E|_{r\sim 0}=h(\rho)(d\psi^2+f_2(\rho)d\rho^2-\sinh^2\rho dt^2)\\
    &h(\rho)=\sqrt{\cosh \rho}~;~f_2(\rho)=(1-\kappa^2 \sinh^2\rho)^{-1}
    \label{e4.7}
\end{align}
which clearly reveals a conformal Krylov basis for $\psi=0$ and $\kappa =0$. On the other hand, for small deformations ($\kappa \ll 1$), one deviates from the Krylov basis. 

In order to describe the trajectory of the particle, we parametrize the curve with the choice $\rho=\rho(t)$ and $\psi=0$. This yields the following action for the massive particle
\begin{align}
\label{e4.8}
    S=\int dt L~;~L=-m \sqrt{h(\rho)}\sqrt{\sinh^2\rho -f_2(\rho)\dot{\rho}^2}.
\end{align}
 Like before, we have a cut-off $\rho_B =\sinh^{-1}(1/\kappa)$ in the bulk and the trajectory of the particle (and hence the action \eqref{e4.8}) is valid only in the domain $\rho<\rho_B$.

The Hamiltonian constraint is given by
\begin{align}
\label{e4.9}
    H=H_0=P_\rho \dot{\rho}-L=\frac{m \sqrt{h(\rho)}\sinh^2\rho}{\sqrt{\sinh^2\rho -f_2(\rho)\dot{\rho}^2}}.
\end{align}

The equation of motion that follows from \eqref{e4.8} can be expressed as
\begin{align}
    &\frac{d}{dt}\Big[ \frac{\sqrt{h(\rho)}f_2(\rho)\dot{\rho}}{\sqrt{\sinh^2\rho -f_2(\rho)\dot{\rho}^2}}\Big]+\frac{1}{2\sqrt{h(\rho)(\sinh^2\rho -f_2(\rho)\dot{\rho}^2)}}\nonumber\\&\Big[\partial_\rho h \sinh^2\rho \nonumber\\
    &+h(\rho) \sinh 2\rho -(\partial_\rho h f_2(\rho)+h(\rho)\partial_\rho f_2)\dot{\rho}^2 \Big]=0
\end{align}
which is equivalent to solving the constraint condition \eqref{e4.9}.

Inverting the constraint eq.\eqref{e4.9}, we obtain 
\begin{align}
\label{e4.11}
    \dot{\rho}^2=\frac{\sinh^2 \rho}{f_2(\rho)}\Big[1-\frac{h(\rho)}{\bar{H}^2_0}\sinh^2\rho \Big]~;~\bar{H}_0=\frac{H_0}{m}.
\end{align}

The above eq.\eqref{e4.11} should be solved with the boundary condition $\dot{\rho}(t=0)\ll 1$. From the constraint condition \eqref{e4.9}, this implies $\bar{H}_0 \sim \sinh \rho_\Lambda \gg 1$ where $\rho_\Lambda<\rho_B$ is the UV cut-off such that $\kappa \sinh \rho_\Lambda <1$. Therefore, exactly at LO, we have an equation
\begin{align}
    \dot{\rho}=\frac{\sinh \rho}{\sqrt{f_2}}=\sinh \rho \sqrt{1-\kappa^2 \sinh^2\rho}+\mathcal{O}(1/\bar{H}^2_0)
\end{align}
where we consider only the absolute value of the radial velocity.

For the undeformed ($\kappa =0$) case, one recovers the results of the Krylov basis \cite{Caputa:2024sux}
\begin{align}
\label{e4.13}
    \tanh \bar{\rho}= \frac{\tanh \rho_\Lambda}{\cosh t}.
\end{align}

Taking a derivative with respect to time ($t$) and thereby setting $t=0$, one should be able to recover one of the boundary conditions, namely $\dot{\rho}(t=0)=0$. In the presence of YB effects ($\kappa \neq 0$), one deviates from the solution \eqref{e4.13} pertaining to a Krylov basis, and instead the solution depends on the deformation parameter ($\kappa$)
\begin{align}
    \rho = \bar{\rho}(t)+\kappa^2 y(t) + \mathcal{O}(\kappa^4).
\end{align}

The equation corresponding to $y(t)$ turns out to be
\begin{align}
    \dot{y}= \coth t y(t)-\frac{1}{2}\sinh^{-3}t
\end{align}
where we set $\tanh \rho_\Lambda \approx 1$, in the limit $\rho_\Lambda \gg 1$.

The corresponding solution has a regular and a divergent piece. The divergent piece has a geometrical origin, which comes from the singularity in the metric \eqref{e4.6}-\eqref{e4.7} near the location of the holographic screen. This can be normalized by adding appropriate counter term in the point particle action \eqref{e4.8}. The finite part, on the other hand, reads as
\begin{align}
    y(t)=c_1 \sinh t-\frac{\cosh t}{3}.
\end{align}
Taking the derivative with respect to time and setting $t=0$ afterwards, we find the initial radial velocity along the geodesic of the particle to be non zero $\dot{y}(0)=c_1$.
%%%%%%%%%%%%%%%%%%%%%%%%%%%%%%%%%%%%%%
\subsection{Proper momentum and Krylov complexity}
The proper radial distance ($\bar{\rho}$) is defined as the distance between two nearest points along the radial direction ($\rho$) of the geodesic for a fixed time $\Delta t=0$ together with $\Delta \psi =0$. 

From \eqref{e4.6}, this implies that
\begin{align}
    ds^2 = d \bar{\rho}^2=h(\rho)f_2 (\rho) d\rho^2 \Rightarrow d\bar{\rho}=\frac{\sqrt{h(\rho)}}{\sqrt{1-\kappa^2 \sinh^2 \rho}}d\rho
\end{align}
which upon integration yields solution in terms of the Appell function of the first kind,
\begin{align}
    \bar{\rho}&=-\frac{\sqrt{2} }{\kappa ^2}\nonumber\\&F_1\Big(\frac{1}{2};\frac{1}{2},\frac{3}{8};\frac{3}{2};\frac{1}{2} \left(-\kappa ^2\cosh (2 \rho ) +\kappa ^2+2\right),\nonumber\\&\frac{-\cosh (2 \rho ) \kappa ^2+\kappa ^2+2}{2 (1+\kappa ^2)}\Big)\nonumber\\
    & \times \sqrt[4]{\cosh \rho } \text{csch}(2 \rho ) \left(\frac{\kappa ^2 \cosh ^2\rho }{\kappa ^2+1}\right)^{3/8} \nonumber\\&\sqrt{\kappa ^2 \sinh ^2\rho  \left(\kappa ^2 (-\cosh (2 \rho ))+\kappa ^2+2\right)}.
\end{align}

Proper momentum is defined as the canonical momentum defined along the proper radial coordinate ($\bar{\rho}$) of the geodesic
\begin{align}
\label{e4.19}
    P_{\bar{\rho}}=\frac{\partial L}{\partial \dot{\bar{\rho}}}=P_\rho \frac{\partial \dot{\rho}}{\partial \dot{\bar{\rho}}}=\frac{m \sqrt{f_2(\rho)}\dot{\rho}}{\sqrt{\sinh^2\rho -f_2(\rho)\dot{\rho}^2}}=\frac{H_0 \sqrt{f_2(\rho)}\dot{\rho}}{\sqrt{h(\rho)}\sinh^2\rho}
\end{align}
where in the last equality, we have used the constraint \eqref{e4.9}.

Expanding \eqref{e4.19} up to $\mathcal{O}(\kappa^2)$, we obtain
\begin{align}
    -P_{\bar{\rho}}/H_0 &= \sinh t \sqrt[8]{\tanh ^2t}-\frac{\kappa^2}{16}\sqrt[8]{\tanh ^2t} \text{csch}t \text{sech}t\nonumber\\ &\times  \left(4 c_1 \sinh ^3t (6 \cosh (2 t)+7)-7 \cosh t-\cosh (5 t)\right)\nonumber\\&+\mathcal{O}(\kappa^4).
\end{align}

An expansion close to $t \sim 0$ reveals a finite contribution as well as a divergent contribution. The divergent contribution has its source in the metric singularity, which can be regularized by adding a counter term, as mentioned before. A closer inspection reveals that this divergence has its origin in the term $\sqrt{f_2}$ in \eqref{e4.19}. Considering the finite contribution, we find up to LO in the expansion in the YB parameter
\begin{align}
    -P_{\bar{\rho}}/H_0 &= t^{5/4}\Big( 1+ \Big(\frac{5}{8}-\frac{13}{4}\frac{\delta y}{\delta t}|_{t=0}\delta t \Big)\kappa^2 \Big)+\cdots
\end{align}
which clearly reveals that the complexity is suppressed due to deformations in space time. Notice that this suppression has its origin in the initial velocity of the particle near the boundary, which is a typical characteristic of YB deformation, as we have witnessed before. A closer comparison with \cite{Caputa:2024sux} reveals that the scaling at an early time (for $\kappa =0$) differs by a factor $t^{1/4}$, which has its origin in the overall scale factor $h^{1/4}(\bar{\rho})$ appearing in \eqref{e4.19}. 
%%%%%%%%%%%%%%%%%%%%%%%%%%%%%%%%%%%%%%%%%%%%%%%%%%
\section{Comments on the dual quantum mechanical model}
\label{sec6}
 Here, we outline the nature of the dual field theory for each of the cases studied in this paper. In the example of the deformed $AdS_2\times S^2 \times T^6$, the background \eqref{e2.8} is conformally equivalent to a CFT$_1$. The conformal factor $\Omega(\rho)$ introduces a finite radial cut-off \eqref{e2.9} along the bulk radial direction ($\rho$). One should think of this dual theory as ($0+1$)d quantum mechanics (QM) with a cut-off, which eventually breaks the underlying scale invariance at short distances. Here, the holographic screen acts like a physical cut-off regulating the UV divergences in the boundary quantum mechanical description; see Fig.\ref{figYB}. 

The QM with a cut-off (that lives on the holographic screen) should be thought of as the low energy ``effective description'' of the quantum degrees of freedom living on the boundary ($\rho_\infty$) of the undeformed AdS$_2$. In the supergravity counterpart, this is achieved by integrating out the bulk degrees of freedom between the boundary ($\rho_\infty$) and the cut-off ($\rho_B$) or equivalently integrating out the degrees of freedom above the UV cut-off scale, $\Lambda_{UV} =\frac{ 1}{\kappa L}$ (where $L$ is the length of the AdS as introduced in \eqref{e2.1}) in the boundary theory. Clearly, in the undeformed ($\kappa \sim 0$) limit, this UV cut-off is pushed towards infinity, and as a result, one recovers the full quantum mechanical description on the boundary. 

In terms of a one dimensional Krylov chain \cite{Caputa:2024sux}, the new cut-off ($\Lambda_{UV}$) should be reflected as a change ($\epsilon \rightarrow \epsilon + \delta \epsilon$) in the lattice spacing between neighboring lattice points $|K_n)$ and $|K_{n+1})$. In other words, $\delta \epsilon$ introduces a change in the lattice cut-off as opposed to the undeformed ($\kappa=0$) theory \cite{Caputa:2024sux}. As we argue below, this modified cut-off ($\delta \epsilon$) should affect the Hamiltonian and the complexity of the low energy effective theory. 

At the level of the one dimensional Krylov chain, the new cut-off ($\Lambda_{UV}$) would increase the lattice spacing ($\delta \epsilon >0$) between adjacent lattice points. Clearly, this modified lattice spacing would correspond to a modified Hamiltonian $H_{eff} \sim H_K + \lambda \bar{H}$ that should satisfy the Krylov chain criterion \cite{Parker:2018yvk},\cite{Balasubramanian:2022tpr} for operator complexity. Here $\lambda =\frac{\delta \epsilon}{\epsilon}$ is a relevant coupling. In summary, we have a new set of Krylov basis states $|K_n)$ corresponding to the effective Hamiltonian ($H_{eff}$), which could be thought of as a new set of lattice points on a one dimensional Krylov chain with modified lattice distance between them.  Clearly, in the true limit $ \frac{\delta \epsilon}{\epsilon} \rightarrow 0$, one should be able to recover the results of the undeformed theory \cite{Caputa:2024sux}. 

Substituting into the Krylov chain condition \cite{Balasubramanian:2022tpr}, one would immediately recognize that the Lanczos coefficients ($b_n$) would have been modified accordingly in order to maintain a consistent Krylov criterion for the effective Hamiltonian, where $\bar{H}$ would satisfy an analog condition. In other words, the Lanczos coefficients ($b_n$) would receive corrections due to finite cut-off ($\Lambda_{UV}$), which has consequences as we elaborate below.

Typically, in the presence of a cut-off, we have a finite dimensional Krylov subspace where the Lanczos coefficients ($b_n$) are sensitive to the UV cut-off \cite{Avdoshkin:2022xuw}. We expect a similar situation to hold for quantum mechanical models dual to YB deformed $AdS_2\times S^2 \times T^6$. At late times, the Lanczos coefficients ($b_n$) would be sensitive to the UV cut-off \cite{Hashimoto:2023swv}, which will stop being linear and thereby exhibit a saturation near the cut-off $\Lambda_{UV}$. The saturation would depend on the underlying lattice model or equivalently on the deformation parameter ($\kappa$) on the supergravity counterpart. This could be tested explicitly by computing the variance $\sigma^2=\langle x^2 \rangle -\langle x \rangle^2$ associated with the Lanczos coefficients ($x_n=\log (b_{2n-1}/b_{2n})$). In a finite dimensional Krylov space, the late time behavior of Krylov operator growth is sensitive to variance $\sigma^2$, while the latter decreases with the increase in the lattice spacing.
%%%%%%%%%%%%%%%%%%%%%%%%%%%%%%%%%%%%%%%%%
\subsection{YB effects on continuum QFTs}
In continuum QFTs, the UV singularity of the two-point correlation leads to a high-frequency exponential tail in the spectral density, which generically yields the asymptotically linear growth of the Lanczos coefficients $b_n \sim n$. On the other hand, any relevant deformation in the field theory can modify the infrared dynamics and the intermediate scale geometry. However, it cannot remove the ``fundamental'' UV degrees of freedom or the qualitative aspects of the underlying short-distance singularity of the continuum theory. In summary, a relevant operator alone cannot generate a genuine physical lattice cut-off and thereby cannot cause strict late-time saturation of the Lanczos coefficients $b_n$. 

From the dual bulk gravity perspective, the Yang Baxter (YB) deformation modifies the asymptotic boundary structure by introducing the "holographic screen" (or radial cut-off) to a finite radial position $\rho_\Lambda$. In this YB framework, the radial cut-off in the bulk serves as an \textit{effective} or \textit{phenomenological} UV regulator for the boundary QFT, that is quite analogous to an effective lattice/discretization scale $\Delta_{\text{UV}}$. In other words, the YB (or the relevant) deformation alone does not intrinsically discretize the QFT. Rather, the presence of the finite holographic screen acts as an ``external'' UV cut-off. To achieve a finite-dimensional Krylov space or complete saturation of $b_n$, an additional explicit UV regulator (such as a lattice discretization on the boundary) is required. The phenomenological cut-off $\Delta_{\text{UV}}$ due to YB deformation modifies the scaling behavior of the Lanczos coefficients $b_n$ at an intermediate scale, causing a deviation or flattening at scales corresponding to the deformation scale $\kappa$, before the asymptotic behavior is probed.

Let us make the above statements more precise. For example, we consider the case of the continuum QFTs with broken scale invariance. The gravitational dual corresponds to the deformed $AdS_3 \times S^3 \times T^4$ solution in the bulk. On the other hand, the dual QFT is conformally equivalent to a $\text{CFT}_2$ deformed by a heavy operator $(\mathcal{O}(t))$ dual to a massive (particle) probe inserted near the boundary (or the holographic screen). 

The conformal factor \eqref{e3.7} introduces an effective UV cut-off (or the phenomenological short distance cut-off) in the boundary description. The complete action can be schematically expressed as
\begin{equation}
\label{104}
S = S_{\text{CFT}_2} + \lambda \int dt \mathcal{O}(t) J(t)
\end{equation}
where $\lambda = \kappa^2$ is a relevant coupling of the boundary theory. As mentioned before, the deformation \eqref{104} alone does not alter the fundamental continuum nature of the QFT at arbitrarily high energies and instead preserves the short-distance singularity in the QFT. The relevant deformation \eqref{104}, induced by the YB effects, modifies the intermediate scale growth of Lanczos coefficients $b_n$, leading to a saturation like suppression near the deformation scale $\kappa$. A genuine finite-dimensional Krylov subspace and exact saturation would require an additional explicit lattice regularization on the boundary.
%%%%%%%%%%%%%%%%%%%%%%%%%%%%%%%%%%%%%%%%%%%%%%%%%%%%%%%%%%
\section{Summary and conclusions}
\label{sec7}
We identify the dual Krylov basis for a class of Yang-Baxter (YB) deformed supergravity backgrounds that have been constructed in recent years. We consider three different examples of such deformed backgrounds containing and AdS$_2$ and AdS$_3$ factor for the first two categories and a mirror dual of the (deformed) AdS$_3$ solution. In the first example of the deformed AdS$_2$ solution, the geometry appears to be conformally equivalent to a Krylov basis. In the second example of deformed AdS$_3$, the Krylov basis is identified at a special point of the parameter space where both left and right deformation parameters appear to be the same, namely $\kappa_L=\kappa_R$. The Krylov basis associated with the mirror dual has been identified following a double Wick rotation and setting the deformation parameter to zero.

The key features of our analysis are the following: (i) We show that the effects of YB deformation is to produce a finite cut-off (the so called holographic screen) in the bulk, which from the perspective of an observer living at the boundary is to measure a non-zero growth of Krylov complexity to begin with. In the the bulk this corresponds to a finite initial velocity of the particle, unlike the case with pure AdS. (ii) We show that the rate of growth of complexity is suppressed with increasing YB deformation, which is an artifact of the reduced size of the Hilbert space associated with the Krylov basis.

Before we conclude, we wish to list down a set of projects that might be worth pursuing.

$\bullet$ It would be interesting to extend the above analysis for YB deformed solutions containing a deformed AdS$_n$, where $n>3$. An ideal example is the Jordanian deformed AdS solution in type IIB \cite{Kawaguchi:2014fca}, which is constructed using $R$-matrices that satisfy the classical YB equation. The complete solution can be characterized in terms of the 10d metric containing a deformed AdS$_5$ factor times a $S^5$, along with the background NS-NS and RR fluxes. Interestingly, the background dilaton ($\Phi$) appears to be zero, which therefore does not play a role in the particle dynamics as one goes from the string frame to the Einstein frame.

$\bullet$ The above results could be further generalized for a class of YB deformed string backgrounds generated through $R$-matrices that contain two or three parameter generalization \cite{Matsumoto:2014ubv}. The three parameter deformed backgrounds are in particular of interest as they can be obtained following a TsT and S duality, starting from the undeformed $AdS_5 \times S^5$.

%%%%%%%%%%%%%%%%%%%%%%%%%%%%%%%%%%%%%%%%%%%%%%%%%%%%%%%%%%%%%%%%%%%%%%
\paragraph*{Acknowledgements.}
 The author would like to thank Carlos Nunez for his comments on the draft. The author acknowledges the Mathematical Research Impact Centric Support (MATRICS) grant (MTR/2023/000005) received from ANRF, India. \\ 
%%%%%%%%%%%%%%%%%%%%%%%%%%%%%%%%%%%%%%%%%%%%%%%%%%%%%%%%%%%%%%%%%%%%%%%%%%%%%%%%%%%

\end{document}